\documentclass[12pt,a4paper]{ijicic}

\def\currentvolume{x}

\def\ppages{0--0}

\usepackage{amsmath}
\usepackage{amssymb}
\usepackage[dvips]{graphicx}
\usepackage{epsfig}
\usepackage{epsf}
\include{figure}
\usepackage{verbatim}

\title{NEURAL FEEDBACK SCHEDULING OF REAL-TIME CONTROL TASKS}

\author[F. Xia, Y.-C. Tian, Y. Sun and J. Dong]{}

\begin{document}
\maketitle

\begin{center}
\normalsize{\scshape Feng Xia$^{1,3}$, Yu-Chu Tian$^{1,*}$, Youxian
Sun$^2$ and Jinxiang Dong$^3$}
\medskip

$^1$Faculty of Information Technology \\
Queensland University of Technology\\
GPO Box 2434, Brisbane QLD 4001, Australia\\
\{f.xia, y.tian\}@qut.edu.au\\
$^*$Corresponding author
\medskip

$^2$State Key Laboratory of Industrial Control Technology\\
Zhejiang University\\
Hangzhou 310027, P. R. China
\medskip

$^3$College of Computer Science and Technology\\
Zhejiang University\\
Hangzhou 310027, P. R. China
\end{center}

\medskip
\centerline{Received August 2007; revised December 2007}
\medskip

\begin{abstract}
{\em Many embedded real-time control systems suffer from resource
constraints and dynamic workload variations. Although optimal feedback
scheduling schemes are in principle capable of maximizing the
overall control performance of multitasking control systems, most of
them induce excessively large computational overheads associated
with the mathematical optimization routines involved and hence are
not directly applicable to practical systems. To optimize the
overall control performance while minimizing the overhead of
feedback scheduling, this paper proposes an efficient feedback
scheduling scheme based on feedforward neural networks. Using the
optimal solutions obtained offline by mathematical optimization
methods, a back-propagation (BP) neural network is designed to adapt
online the sampling periods of concurrent control tasks with respect
to changes in computing resource availability. Numerical simulation
results show that the proposed scheme can reduce the computational
overhead significantly while delivering almost the same overall
control performance as compared to optimal feedback scheduling.}
\end{abstract}

\section{Introduction}
Embedded control systems have been used in a wide variety of
applications. These systems are typically resource
constrained due to various technical and economic reasons
\cite{1,2,3,4,5}. In particular, the computing speeds of most
embedded processors are limited as compared to general-purpose
computers. Also, it is common that multiple control tasks
have to compete for the use of one processor. For a real-time
embedded control system, such resource constraint may
affect the system timing behaviour significantly and may even yield
unsatisfactory control performance. This problem will be further
pronounced when the system operates in dynamic environments where
the CPU workload varies over time. In the context of resource
constraints, these dynamic variations in workload will possibly lead
to low CPU utilization and/or system overloading. As a consequence, the
performance of a multitasking control system will be jeopardized
\cite{4,6}.

Recently, feedback scheduling \cite{1,4,7} has emerged as a
promising technology for addressing the above mentioned uncertainty
in resource availability. The basic idea of feedback scheduling is
to allocate available resources dynamically among multiple real-time
tasks based on feedback information about actual resource usage. In
multitasking control systems, a straightforward objective of
feedback scheduling is to optimize the overall quality of control
(QoC) characterized by some sort of performance indices. Accordingly, the problem of feedback scheduling can be
formulated as a constrained optimization problem, which is usually
referred to as optimal feedback scheduling \cite{1}. In this
optimization problem, the total control cost is to be minimized
through optimizing scheduling parameters of control tasks under the
constraint of system schedulability. The most popular solution for
this optimization problem is based on mathematical optimization
algorithms, e.g., \cite{8,9,10,11,12}. Since feedback schedulers are usually
executed at runtime, it is of paramount importance to take into
account the computational overhead of the scheduling
algorithm to be employed \cite{10,11}. If the feedback scheduler
consumes too much computing resources, the execution of control
tasks will inevitably be impacted in the presence of resource
constraint. This may then cause significant degradation of the
overall QoC. In theory optimal feedback scheduling schemes
are effective in optimizing the overall QoC, but optimization solutions
typically involve complex computations, which induce large overheads. Therefore, they are not suitable for online use in most cases.

To tackle the problem associated with the large computational
overheads of optimal feedback scheduling algorithms, a neural
feedback scheduling (NFS) scheme will be proposed in this paper. The
goal is to optimize the overall QoC of multitasking control systems
through feedback scheduling while minimizing the scheduling
overhead. A feedforward back-propagation (BP) neural network with a simple structure is adopted to build the feedback scheduler. The
scheme has the advantages of low computational overhead, wide
applicability, and intelligent computation. It can also
deliver almost optimal QoC.

Much effort has been made to approximate the optimization solutions
using simpler algorithms that incur smaller overheads. Cervin
\emph{et al}. \cite{13} presented a linear proportional rescaling
method. Castane \emph{et al}. \cite{11} developed a heuristic
approximation of the optimization procedure. However, these
approaches are more or less ad hoc. For example, the rescaling
method in \cite{13} is mainly intended for systems with
approximately linear or quadratic cost functions. In contrast, NFS
is based on a formal and well-established technology, i.e., neural
networks, and is consequently widely applicable. While neural
networks have proven to be a highly effective technology for solving
scientific and engineering problems \cite{14,15,16,17}, the
application of neural networks in feedback scheduling remains
unexplored.

The rest of this paper is organized as follows. Section 2 formulates
the problem of optimal feedback scheduling. The neural feedback scheduling scheme is proposed in
Section 3. Section 4 evaluates the performance of the proposed scheme via
numerical simulations. Finally, the paper is concluded in Section 5.

\section{Problem Formulation}
Consider a system where $N$ independent control tasks running on a
processor with limited processing capability. In addition to control
tasks, other non-control tasks with higher priorities may run
concurrently. The timing attributes of these non-control tasks
cannot be manipulated intentionally. The execution times of control
tasks and the requested CPU utilization of non-control tasks may
change over time. The feedback scheduler adapts the sampling periods
of the control tasks to workload variations so that the CPU
utilization is maintained at a desired level. For simplicity, assume
that all task execution times and the CPU workload are available at
run-time.

According to sampled-data control theory, smaller sampling periods
yield better control performance. However, the decrease in sampling
period will result in an increase in the requested CPU utilization
of the relevant control task. In extreme cases the schedulability of
the system may be violated, and hence the control performance will
deteriorate due to deadline misses. In order to optimize the overall
QoC, the sampling periods should be adjusted under the constraint of
system schedulability \cite{18}. From the optimization point of view,
the available computing resource should be distributed among control
tasks in an optimal way that the total control cost of the system is
minimized.

Let $h_i$ and $c_i$ denote the period and the execution
time of control task $i$, respectively. Optimal feedback scheduling
can be formulated as a constrained optimization problem:

\begin{equation} \label{equ:1}
\begin{split}
\min_{h_1,\cdots,h_N} J=\sum^N_{i=1}w_iJ_i(h_i)\\
\mbox{s.t.}\ \ \
\sum^N_{i=1}c_i/h_i\leq U_R
 \end{split}
 \end{equation}
where $J_i(h_i)$ is the control cost function of loop $i$, as a
function of the period $h_i$; $w_i$ is a weight reflecting the
relative importance of each loop; $U_R$ is the maximum allowable
utilization of all control tasks and is related to the underlying
scheduling policy and the requested utilization of
disturbing tasks. In general, $J_i(h_i)$ is monotonically
increasing.

To obtain linear constraints, the costs are often recast as
functions of sampling frequencies $f_i = 1/h_i$ \cite{8,13}. By argument substitution, (\ref{equ:1}) can be
rewritten as:
\begin{equation} \label{equ:2}
\begin{split}
\min_{f_1,\cdots,f_N} J=\sum^N_{i=1}w_iJ_i(f_i)\\
\mbox{s.t.}\ \ \
\sum^N_{i=1}c_i/f_i\leq U_R
 \end{split}
 \end{equation}
In the above formulation, it is crucial to choose an appropriate
cost function. Linear quadratic cost functions are used in
\cite{8,9,10}. Alternatively, approximate cost functions may also be
used, as in \cite{12,13}. The neural feedback scheduling scheme to
be developed in this paper does not rely on control cost functions
of any specific forms. It is applicable to control systems with
arbitrary cost functions provided that (\ref{equ:2}) can be solved
offline.

In the area of optimization, there exist many well-established
methods for solving the constrained optimization problem formulated
by (\ref{equ:2}). The necessary and sufficient condition for the
optimal solutions is given by Kuhn-Tucker condition \cite{19}, given
that the $J_i(f_i)$ is convex. When $J_i(f_i)$ is not convex, the
Kuhn-Tucker condition becomes a necessary condition. Since
$J_i(f_i)$ is convex for most control systems \cite{8}, the
Kuhn-Tucker condition can be regarded as a general tool for
obtaining the optimal sampling frequencies/periods. Sequential
quadratic programming (SQP) has been recognized as one of the most
efficient methods for solving constrained optimization problems
\cite{20}. For the sake of simplicity, it is assumed hereafter that
the SQP method is by default used for the optimal feedback
scheduling scheme wherever it is involved.

Complex computations associated with gradients and Hessian matrices
will be involved when solving (\ref{equ:2}) using mathematical
optimization algorithms; and a large number of iterations are
usually required before reaching the final solution. This results in high
computational complexity of the algorithms.
In the SQP method, for instance, one or two quadratic programming
sub-problems must be solved in each iteration and thus take a great
deal of time to complete. When applied to feedback scheduling, this
algorithm may introduce significant overheads. In fact, most of
existing optimal feedback scheduling algorithms suffer from the
problem of too large computational overheads, which impair their
practicability.

\section {Neural Feedback Scheduling}
\label{section:Neural Feedback Scheduling}

Since optimal feedback scheduling schemes are generally too
computationally expensive to be used online, schemes with much less
computational complexity are needed. In this section, an efficient
feedback scheduling scheme using neural networks will be proposed.
Some reasons for using neural networks are:

\begin{itemize}
\item Feedforward neural networks with simple structures can yield much smaller
feedback scheduling overheads than mathematical optimization methods;

\item     With regard to the accuracy of the solutions, mathematical optimization
methods generate the accurate optimal solutions offline, which can
be exploited in the design of online feedback schedulers. On the
other hand, neural networks are powerful in learning and adapting,
and are capable of approximating complex nonlinear functions with
arbitrary precision \cite{14,21}. Once well trained using the
accurate optimal solutions at design time, neural networks will be
able to deliver almost-optimal feedback scheduling
performance at runtime; and

\item     The generalization capability of neural networks is also very good in that
they can easily handle untrained input data, noise, incomplete data,
etc. This helps improve the robustness and fault-tolerance of the
feedback scheduler.
\end{itemize}

\subsection{Design methodology}
\label{subsection:Design methodology}

The basic idea behind neural feedback scheduling is to use a feedforward neural network to
approximate the optimal solutions, which are obtained using
mathematical optimization methods. Following this idea, training and
testing data will not be a problem since it can be easily created
offline by applying, say the SQP method, to the optimal feedback
scheduling problem. In the following the structure of neural
feedback scheduler and the design flow will be described.

This paper uses a three-layer feedforward BP network to build the
feedback scheduler. Two major reasons for the choice of a BP network
are as follows:
\begin{itemize}
\item The structure of BP neural networks is simple, which is beneficial
to simplifying online computations; and
\item Due to their somple structure, BP networks are easy to implement, and
are the most widely used neural network technology in practice.
\end{itemize}

\begin{figure}[htbp!]
\centering
\includegraphics[scale=1, bb=114 372 464 504]{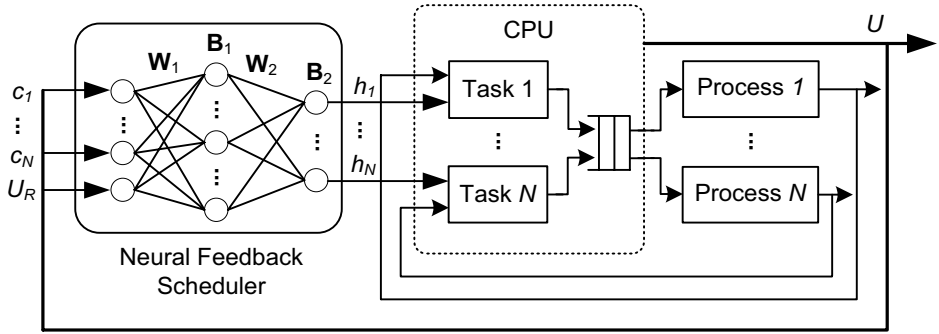}
\caption{\label{fig:Architecture}Architecture of neural feedback
scheduling}
\end{figure}

As shown in Figure \ref{fig:Architecture}, there is only one hidden
layer apart from the input and output layers in the BP network used
as the neural feedback scheduler. Since feedforward neural networks
with only one hidden layer are able to approximate arbitrary
functions with arbitrary precision that are continuous on closed
intervals \cite{14}, one hidden layer is sufficient for guaranteeing
solution accuracy.

According to (\ref{equ:2}), with given cost functions, the values of
sampling frequencies will depend on the execution time $c_i$ of each
control task and the desired CPU utilization $U_R$. As a consequence
of this observation, ($N+1$) inputs, i.e., $c_1, \cdots, c_N$,
$U_R$, are set for the neural feedback scheduler. Since the role of
the feedback scheduler is to determine sampling periods of all
loops, the sampling periods $(h_1, \cdots, h_N)$ or frequencies
$(f_1, \cdots, f_N)$ are natural outputs of  the neural feedback
scheduler.

From a real-time scheduling perspective, both the inputs and outputs
of the feedback scheduler are related to resource utilization. From
a control perspective, sampling periods/frequencies are important
design parameters of the control loops. Therefore, the neural
feedback scheduler establishes a mapping from temporal parameters
(for real-time scheduling) to controller parameters (for real-time
control).

The relationship between the outputs and inputs of the neural
feedback scheduler is expressed as:
\begin{equation}
\label{equ:3}
\textbf{Y}=\textbf{W}_2(\sigma(\textbf{W}_1\textbf{X}+\textbf{B}_1))+\textbf{B}_2
\end{equation}
where \textbf{W}$_1$, \textbf{W}$_2$, \textbf{B}$_1$ and
\textbf{B}$_2$ are weight matrices and bias vectors, respectively;
the input vector \textbf{X}$=[c_1, \cdots, c_N, U_R]^T$; and the
output vector \textbf{Y}$=[f_1, \cdots, f_N]^T$ (or $[h_1, \cdots,
h_N]^T$). The activation functions used are the sigmoid transfer
function $\sigma(x)=\frac{1}{1+e^{-x}}$ in the hidden layer and the
linear transfer function in the output layer.

In designing a neural feedback scheduler, it is of great importance
to determine an appropriate number of neurons in the hidden layer.
Although there are some guidelines for neural network design, there
is no general theory for determining the number of hidden neurons.
Therefore, practical experience and simulation studies must be
relied on in most cases. As Equation (\ref{equ:3}) indicates, the
number of hidden neurons is closely related to the computational
complexity of the neural feedback scheduler. In the case of too many
hidden neurons, the feedback scheduler will consume too much
computing resource, thus causing large feedback scheduling
overheads. Fortunately, the number of control tasks that run
concurrently on the same CPU is typically limited, e.g., less than
10 in most cases \cite{8,9,10,11,12,13}. Therefore, it is usually
unnecessary for the number of hidden neurons to be very large. This
ensures that the computations associated with the neural feedback
scheduler  will not be overly time-consuming. In the training of the
BP network, the Levengerg-Marquardt (LM) algorithm is adopted.

The design flow of the neural feedback scheduler is as follows.
Firstly, formulate the problem in the form of constrained
optimization as given in (\ref{equ:2}). Determine the form of the
cost functions based on control systems analysis, and initialize
related parameters. Secondly, analyze the characteristics of the
execution times of the control tasks to obtain the ranges of their
values. Within the ranges of $c_i$ as well as $U_R$, select a number
of data pairs, and for each pair, use the SQP method to solve the
optimal feedback scheduling problem offline, producing sufficient
sample data sets. Thirdly, determine the number of hidden neurons
according to the number of control loops, and initialize the neural
network. Finally, train and test the neural network using
pre-processed sample data sets. Once the BP network passes the test,
it can thereafter be used online as the neural feedback scheduler.

The online application of the well-designed neural feedback
scheduler is straightforward. At every invocation instant, the
feedback scheduler gathers the current values of all input
variables, and then calculates the sampling periods/frequencies
using (3), followed by the update of sampling period of each loop.

\subsection{Complexity analysis}
The computational complexity of the neural feedback scheduler
proposed above is analyzed below. Though the application of neural
feedback schedulers involves not only online computations but also
offline computations, e.g., mathematical optimization, network
training and test, etc., only the complexity of online computations
is of concern in this paper. This is because it is the computational operations at
runtime that decide the feedback scheduling overhead.

With a given invocation interval, the amount of computing resource
consumed by the feedback scheduler depends directly on the CPU time
needed for each run. To analyze the computational complexity of the
neural feedback scheduler, consider the online calculation
(\ref{equ:3}). Let $M$ be the number of hidden neurons. Equation
(\ref{equ:3}) can be easily decomposed into the following three
sub-equations:
\begin{equation}
\label{equ:4}
\left\{ \begin{aligned}
         \textbf{A} &= \textbf{W}_1\textbf{X}+\textbf{B}_1 \\
         \textbf{Z} &=\sigma(\textbf{A}) \\
         \textbf{Y} &=\textbf{W}_2(\textbf{Z})+\textbf{B}_2
\end{aligned} \right.
\end{equation}
where \textbf{A} $= [a_1, \cdots, a_M]^T$ and \textbf{Z} $= [z_1,
\cdots, z_M]^T$. By substituting
\begin{gather*}
\textbf{W}_1=\begin{bmatrix}w'_{1,1}&\cdots&w'_{1,N+1}\\
\vdots&\ddots&\vdots\\w'_{M,1}&\cdots&w'_{M,N+1}\end{bmatrix},
\textbf{W}_2=\begin{bmatrix}w''_{1,1}&\cdots&w''_{1,M}\\
\vdots&\ddots&\vdots\\w''_{N,1}&\cdots&w''_{N,M}\end{bmatrix},\\
\textbf{B}_1=[b'_1, \cdots, b'_M]^T, \text{and}\
\textbf{B}_2=[b''_1, \cdots, b''_N]^T
\end{gather*}
Equation (\ref{equ:4}) can be rewritten as:
\begin{gather}\label{equ:5}
\begin{bmatrix}a_1\\\vdots\\a_{M-1}\\a_M\end{bmatrix}=
\begin{bmatrix}w'_{1,1}&\cdots&w'_{1,N}&w'_{1,N+1}\\
\vdots&\ddots&\vdots&\vdots\\w'_{M-1,1}&\cdots&w'_{M-1,N}&w'_{M-1,N+1}\\w'_{M,1}&\cdots&w'_{M,N}&w'_{M,N+1}\end{bmatrix}
\begin{bmatrix}c_1\\\vdots\\c_{N}\\U_R\end{bmatrix}+
\begin{bmatrix}b'_1\\\vdots\\b'_N\\b'_{N+1}\end{bmatrix}
\end{gather}
\begin{gather}\label{equ:6}
z_i=\sigma(a_i)=\frac{1}{1+e^{-a_i}},\ \ i=1, \cdots, M
\end{gather}
and
\begin{gather}\label{equ:7}
\begin{bmatrix}f_1\\\vdots\\f_N\end{bmatrix}=
\begin{bmatrix}w''_{1,1}&\cdots&w''_{1,M}\\
\vdots&\ddots&\vdots\\w''_{N,1}&\cdots&w''_{N,M}\end{bmatrix}
\begin{bmatrix}z_1\\\vdots\\z_{M}\end{bmatrix}+
\begin{bmatrix}b''_1\\\vdots\\b''_N\end{bmatrix}
\end{gather}

The above three equations (\ref{equ:5}) through (\ref{equ:7}) give almost all operations that the neural
feedback scheduler has to complete each time when it is invoked. There
are altogether $(4MN+6M-N)$ basic operations associated with these
computations. Clearly, the feedback scheduling overhead relates
primarily to the number of control loops and the number of hidden
neurons, i.e., $N$ and $M$. In general cases, $M$ is proportional to
$N$, i.e., $M\varpropto N$, e.g., $M \approx 2N$. Therefore, the
time complexity of the neural feedback scheduling algorithm is
$O(N^2)$. In contrast, the computational complexity of a typical
mathematical optimization algorithm, e.g., SQP, is (at least)
$O(N^3)$ \cite{11}. Therefore, the neural feedback scheduling can
significantly reduce the computational complexity of the algorithm
in comparison with optimal feedback scheduling.

As mentioned above, the value of $N$ is always limited in real
systems. It will never approach infinity. Consequently, a more
convincing method for examining the runtime efficiency of feedback
schedulers is to compare the actual CPU time consumed by different
feedback schedulers via simulations and/or real experiments, see the
next Section.

\section{Numerical Example}

This section will test and analyze the performance of the proposed
scheme via numerical simulations using Matlab/TrueTime \cite{22}.
From the control perspective, the purpose of this evaluation is
twofold. The first is to validate the effectiveness of the neural
feedback scheduling, i.e., to check whether or not it is able to
deal with dynamic variations in both the control tasks' resource
demands and the available resources. The second is to study the
difference between neural feedback scheduling and \emph{ideal}
optimal feedback scheduling in optimizing the overall QoC. From the
viewpoint of implementation efficiency, the actual time overheads of
different feedback schedulers will be compared, thus highlighting
the major merit of the proposed scheme.

\subsection{Setup overview}
Consider an embedded processor that is responsible for controlling
three inverted pendulums concurrently. Thus, there are three
independent control tasks. The linearized state-space models of the
inverted pendulums are in the form \cite{13}:
\begin{equation}
\label{equ:8}
\begin{aligned}
    \dot{x}(t) &= \begin{bmatrix}0&1\\\omega^2_0&0\end{bmatrix}x(t)+
\begin{bmatrix}0\\\omega^2_0\end{bmatrix}u(t)+v(t) \\
    y(t)&= \begin{bmatrix}1&0\end{bmatrix}x(t)+e(t)
\end{aligned}
\end{equation}
where $\omega_0$ is the natural frequency of the inverted pendulum,
$v$ and $e$ are sequences of white Gaussian noise with zero mean and
variances of $1/\omega_0$ and $10^{-4}$, respectively.

Due to the difference in length, the three inverted pendulums have
different natural frequencies given by $\omega_0$ = 10, 13.3, and
16.6, respectively. All initial states are zero. Every pendulum is
controlled independently by a linear quadratic Gaussian (LQG)
controller, whose objective is to minimize the following cost
function:
\begin{equation} \label{equ:9}
J=\int^\infty_0(y^2+u^2)dt
 \end{equation}
For the sake of simplicity, the approximate cost function given in
\cite{13} is used in (\ref{equ:2}):
\begin{equation} \label{equ:10}
J_i(f_i)=\alpha_i+\gamma_i/f_i
 \end{equation}
where $\gamma_i$ = 43, 67, and 95 for control
loop $i=1, 2, 3$, respective;y. The initial sampling frequency of each loop is chosen as $f_0$
= 58.8, 71.4, and 83.3 Hz, respectively. Also, assume $w_i$ = 1 for
simplicity.

In addition to these three control tasks, there is a periodic
non-control task. The execution time of this task is variable,
causing $U_R$ to vary over time. When the execution of the
feedback scheduling task is neglected, the desired total CPU utilization of all
tasks is set to $\tilde{U}_R = 0.75 < 4(2^{1/4}-1) = 0.76$.
According to \cite{23}, the system schedulability under the rate
monotonic (RM) scheduling policy is guaranteed by $\tilde{U}_R$. The
execution time of the non-control task is $c_4$, and its period $h_4
= 10\ ms$. Therefore, $U_R = \tilde{U}_R - c_4/h_4$, implying that
$U_R$ will change with $c_4$.

Task priorities are assigned as follows. The feedback
scheduling task has the highest priority, and the priorities of
other tasks are determined in accordance with the RM policy. The
invocation interval of the feedback scheduler is $T_{FS}  = 400\ ms$. To
measure the overall QoC of the syste,, the total control cost
$J_{SUM}$ of three control loops is recorded:
\begin{equation} \label{equ:11}
J_{SUM}(t)=\sum^3_{i=1}w_iJ_i(t)=\sum^3_{i=1}\int^t_0(y^2(\tau)+u^2{\tau})d\tau
 \end{equation}

\subsection{Neural feedback scheduler design}
The neural feedback scheduler is designed following the procedures
described in Section \ref{subsection:Design methodology}. Based on
the above description of the simulated system, the following
formulation of the corresponding optimal feedback scheduling problem
is obtained from (\ref{equ:2}).
\begin{equation} \label{equ:12}
\begin{split}
&\min_{f_1,f_2,f_3}J=\alpha+\frac{43}{f_1}+\frac{67}{f_2}+\frac{95}{f_3}\\
&s.t.\ c_1f_1+c_2f_2+c_3f_3\leq 0.75-\frac{c_4}{0.01} \end{split}
\end{equation}
where $\alpha=\alpha_1+\alpha_2+\alpha_3$ is a constant.

For the purpose of creating sample data, the ranges of $c_1$, $c_2$,
and $c_3$ are set to [2, 9], [2, 7], and [1, 7], respectively, with
increments of 1. $c_4$ takes on values ranging from 0.5 to 3 with
increments of 0.5. The units of these parameters are ms. For all
possible values of these parameters, applying the SQP method to
solve (\ref{equ:12}) offline results in 2016 sets of sample data in
total.

To further simplify online computations, the sampling periods
instead of the frequencies are used as the outputs of the neural
feedback scheduler. Once the sample data sets are created, they will
be normalized onto the interval [0, 1]. Since $c_i$ and $U_R$ are on
rather different orders, normalizing original sample data can avoid
saturations of neurons and speed up the convergence of the neural
network. It is not imperative in this work that the sample data be
normalized, because all original data falls inside the interval [0,
1]. In general cases, however, normalization helps improve the
performance of neural networks.

In order to determine the number of hidden neurons, i.e., the value
of $M$, neural networks of different sizes are tested. Given that
the performance is comparable, a smaller $M$ value should be chosen
in order to reduce the feedback scheduling overhead. From this
insight, set $M = 8$ because of the good performance of
corresponding neural network. Once passing the test, the parameters
of the neural network are stored for online use.

\subsection{Results and analysis} 
\begin{figure}[htbp!]
\begin{minipage}[t]{0.47\textwidth} \centering
\includegraphics[scale=0.5, bb=111 269 473 566]{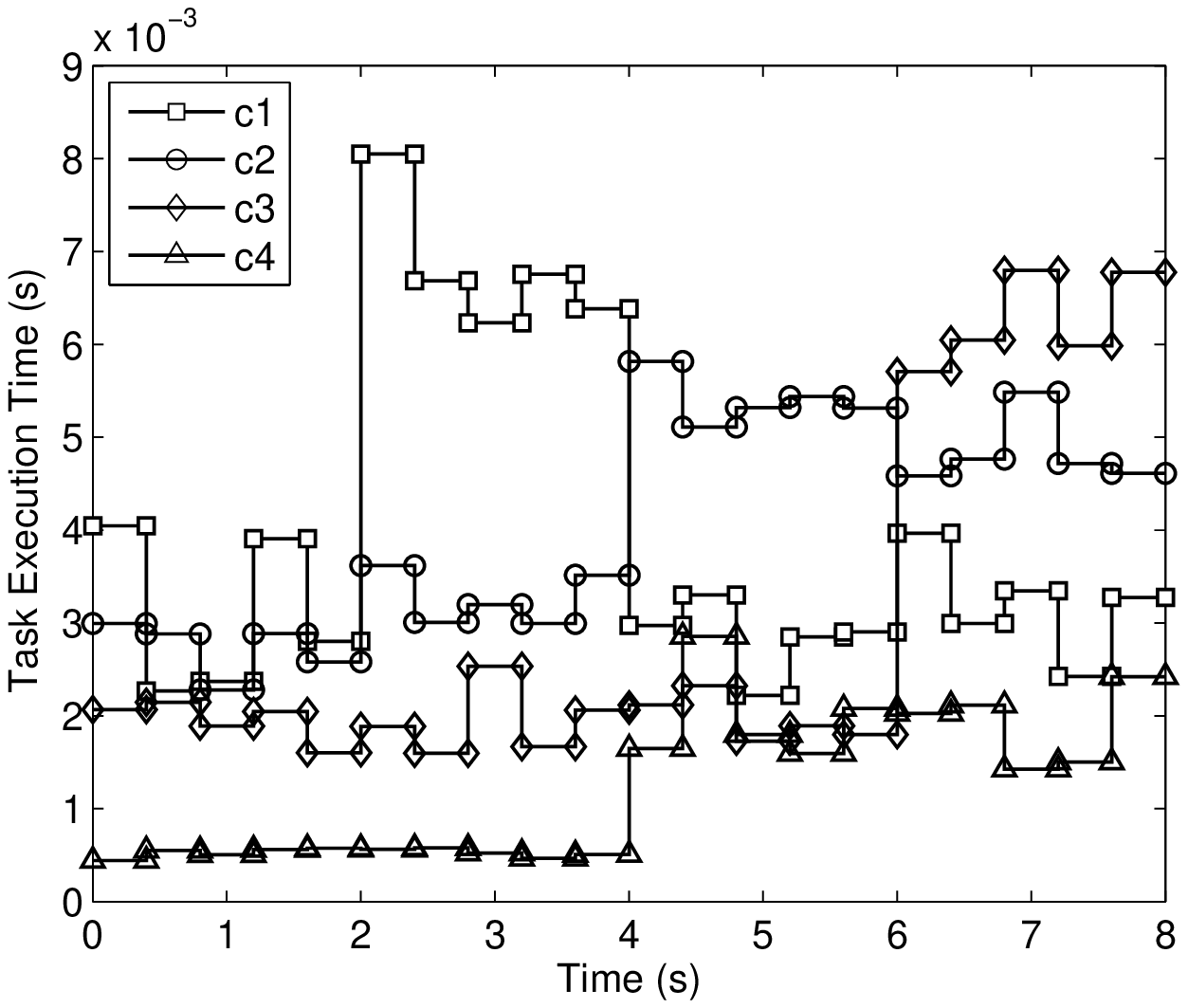}
\caption{Task execution times}
\label{fig:2}
\end{minipage}%
\begin{minipage}[t]{0.47\textwidth}
\includegraphics[scale=0.5, bb=100 267 472
558]{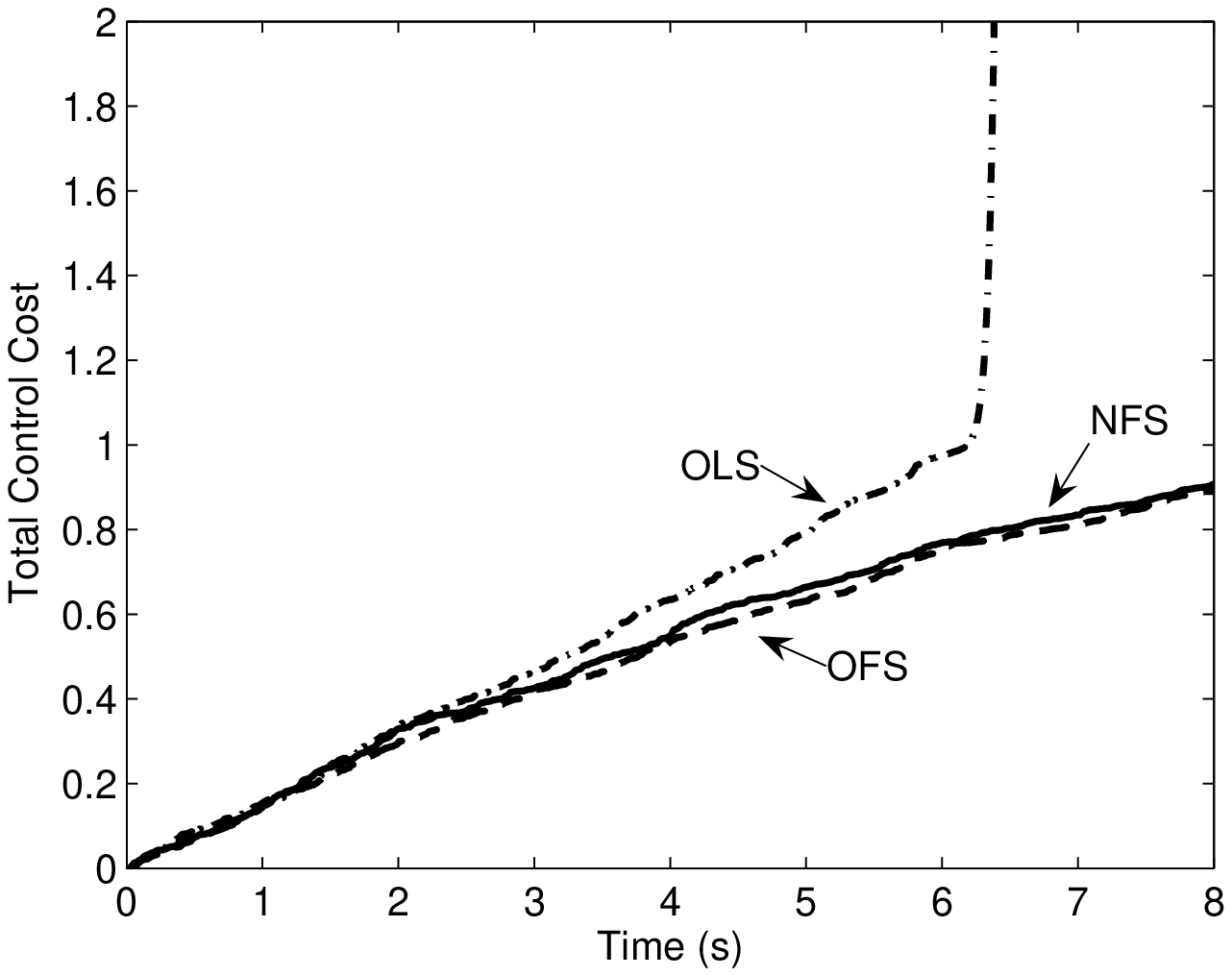} \caption{Total control costs}\label{fig:3}
\end{minipage}
\end{figure}

Let us examine the overall performance
of the control system first. The execution time of each task varies at
runtime according to Figure \ref{fig:2}. The overhead of the feedback
scheduling is neglected here and will be studied later. The
following three schemes are compared:
\begin{itemize}
\item Open-loop scheduling (OLS): All control loops use fixed sampling periods;

\item Optimal feedback scheduling (OFS): The optimal feedback scheduling scheme
that uses the SQP method. This is an idealized case for control performance optimization
 because the online computational overhead is assumed to be zero; and

 \item Neural feedback scheduling (NFS): The method presented in this paper.
\end{itemize}

\begin{figure}[htbp!]
\begin{minipage}[t]{0.47\textwidth}
\includegraphics[scale=0.5, bb=100 267 472
560]{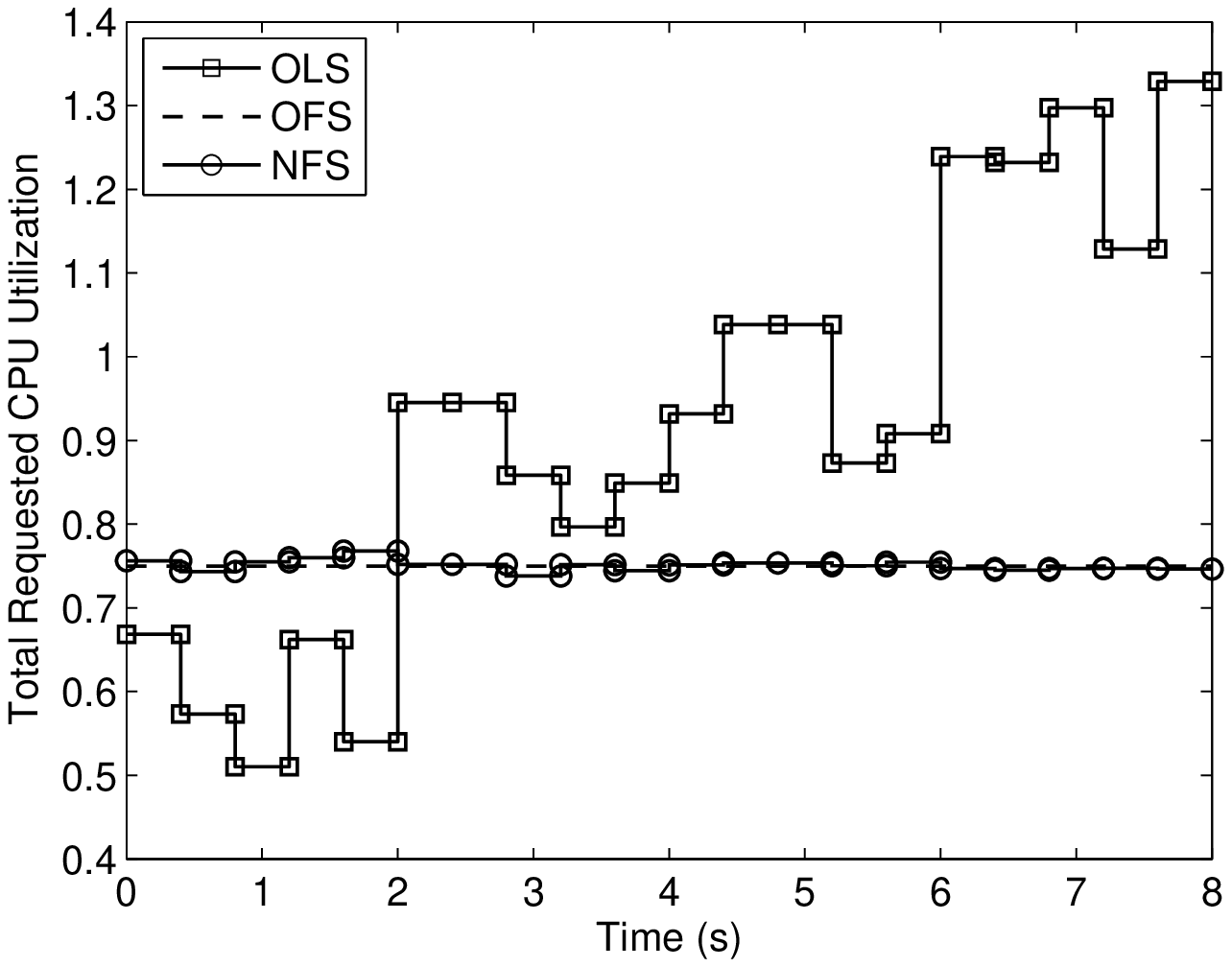} \caption{Total requested CPU
utilization}\label{fig:5}
\end{minipage}%
\begin{minipage}[t]{0.47\textwidth}
\centering
\includegraphics[scale=0.5, bb=98 269 473 566]{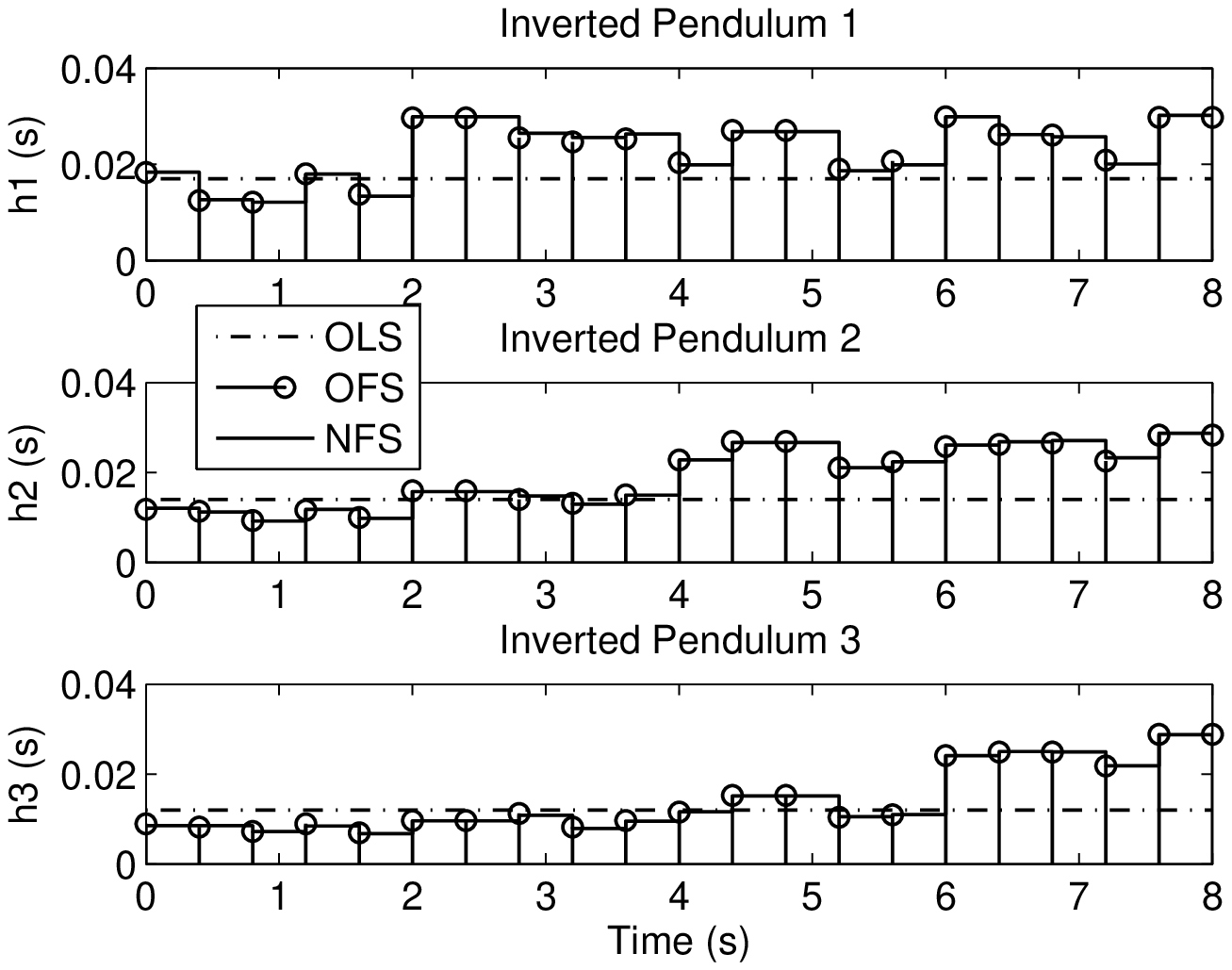}
\caption{Sampling periods}\label{fig:4}
\end{minipage}%
\end{figure}

Figure \ref{fig:3} depicts the total control cost of the system
calculated using (\ref{equ:11}). With the traditional open-loop
scheduling scheme, the system finally becomes unstable. Compared
with the other two feedback scheduling schemes, the open-loop
scheduling yields the worst overall QoC. After time instant $t = 6$s,
at least one pendulum falls down under open-loop scheduling. At $t =
6$s, the task execution times $c_i$ = 0.004, 0.0046, 0.0057, and
0.002s, respectively. The total requested CPU utilization of all
tasks is $U_{req} = \sum(c_i/h_i) = 0.004/0.017 + 0.0046/0.014 +
0.0057/0.012 + 0.002/0.01 = 1.24 > 1$. Obviously, the system is not
schedulable. After this time instant, the total requested CPU
utilization remains very high all along (see Figure \ref{fig:5}),
thereby leading to system instability. Furthermore, according to the
principle of the RM algorithm, the priority of task 1 is the lowest due
to its largest period. Therefore, the first pendulum is finally out
of control.

By comparing NFS with OLS via Figure \ref{fig:3}, it is found that
NFS is effective in dealing with dynamic variations of both task
execution times and available resources. The comparison of NFS and
OFS, on the other hand, indicates that NFS can deliver almost the
same overall QoC as OFS.

To examine the difference between NFS and OFS in more detail, Figure
\ref{fig:4} depicts the sampling periods of three control loops
under different schemes. In contrast to the fixed sampling periods
under OLS, both NFS and OFS adapt sampling periods at runtime. All
sampling periods under NFS and OFS are nearly the same, again indicating that neural feedback scheduling  delivers almost the same results as optimal feedback scheduling.

The adjustment of sampling periods results in changes in the total
requested CPU utilization of all tasks. As shown in Figure
\ref{fig:5}, when the open-loop scheduling scheme is used, the CPU
workload changes with task execution times, because all task periods
are fixed. After time instant $t = 6$s, the total requested CPU
utilization is always higher than 100\%, thereby incurring severe
overload conditions. On the contrary, OFS and NFS are able to keep
the (requested) CPU utilization at or very close to the desired
level $\tilde{U}_R$ = 75\% by means of dynamic adjustment of task
periods. The system schedulability is therefore always guaranteed
under OFS and NFS.

In the following, the runtime overhead of the neural feedback
scheduler is examined, in comparison with the optimal feedback
scheduler. Both feedback schedulers are implemented in the same
environment using Matlab. The hardware platform is the same PC
running Microsoft Windows XP. This environment cannot provide
real-time guarantees. It is used here only for the comparison of
different feedback schedulers in terms of computational overhead.
Attention should be paid to the relative values rather than the
absolute values of the CPU time the feedback schedulers consume.

For the optimal feedback scheduler and the neural feedback scheduler
designed above, the CPU time they actually expend for 500
consecutive runs is recorded, respectively. In each run, task
execution times are randomly drawn from the sets given in Figure
\ref{fig:2}.

\begin{figure}[htbp]
\begin{minipage}[t]{0.5\textwidth} \centering
\includegraphics[scale=0.5, bb=95 271 482 573]{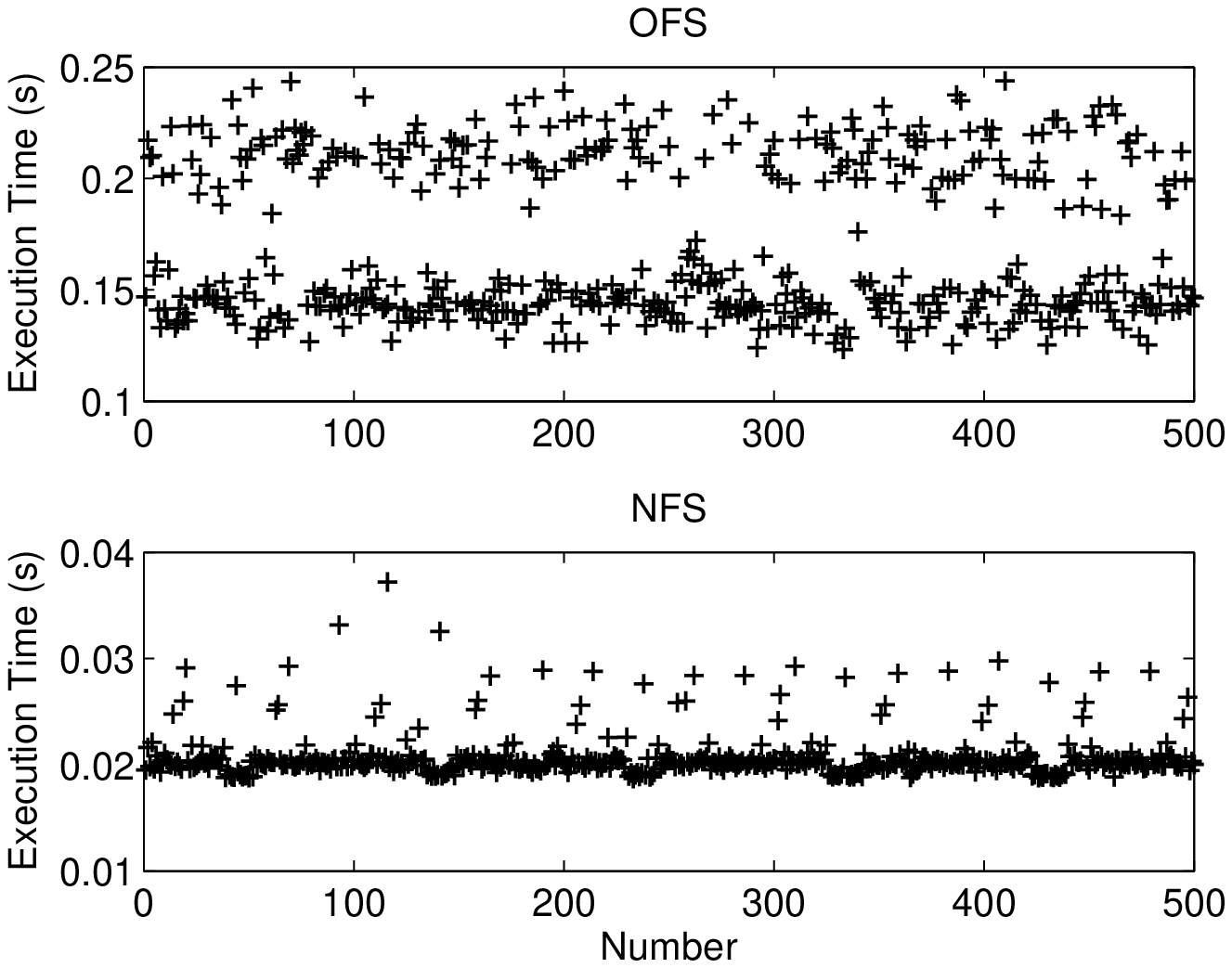}
\caption{Recorded execution times of feedback
schedulers}\label{fig:6}
\end{minipage}%
\begin{minipage}[t]{0.55\textwidth}
\includegraphics[scale=0.5, bb=48 283 410
556]{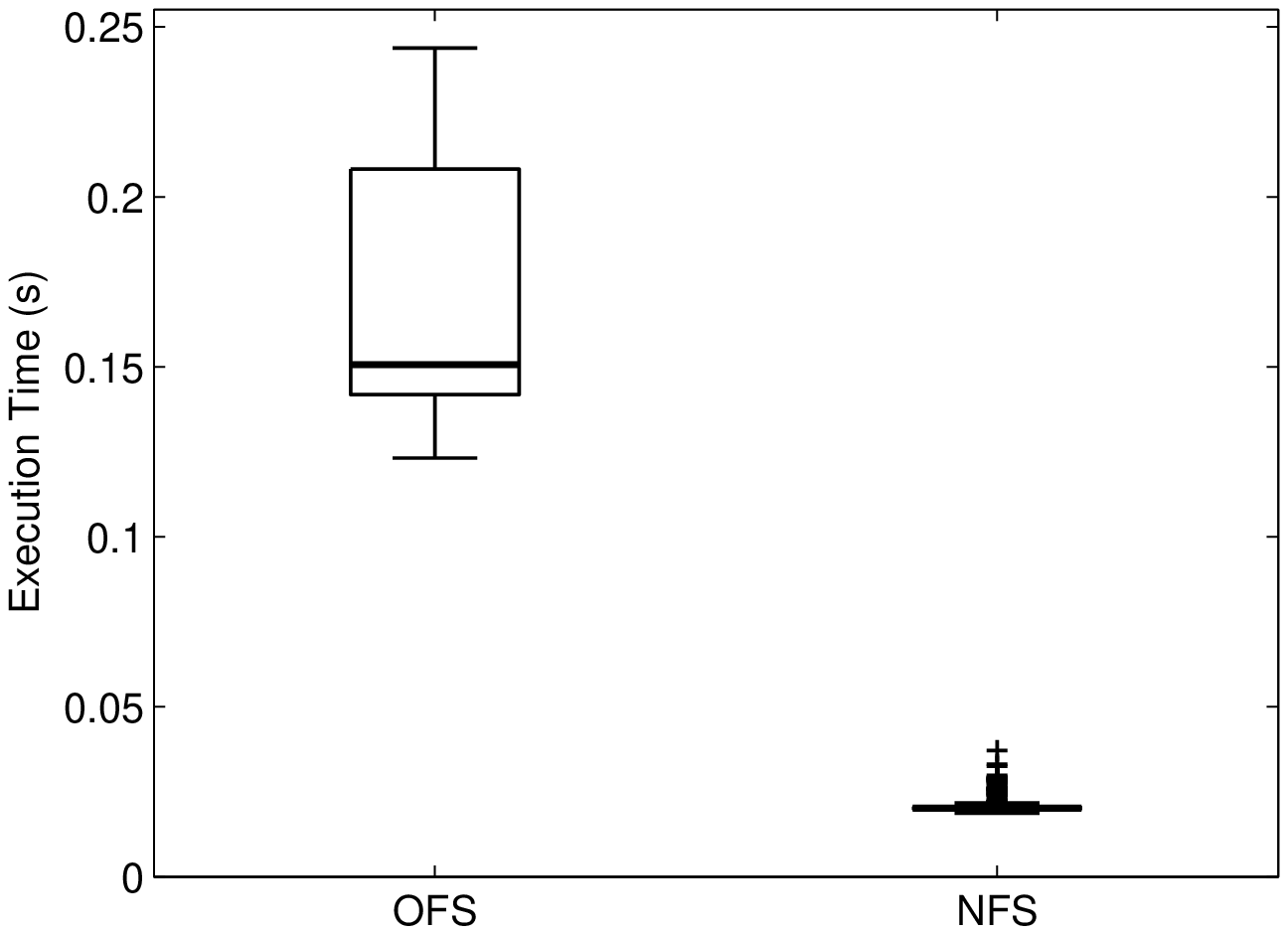}
\caption{Box and whisker plot for execution
times of feedback schedulers}\label{fig:7}
\end{minipage}
\end{figure}

As shown in Figure \ref{fig:6}, the execution time of the optimal
feedback scheduler using SQP falls between 0.12s and 0.25s in most
cases, with the average of 0.1701s. The execution time of the neural
feedback scheduler is always less than 0.04s, and close to 0.02s in
most cases. The average becomes 0.0207s. The ratio of time
overhead of OFS to NFS approximates 0.1701:0.0207 $\approx$ 8.22:1.
The overhead of NFS is only 12.2\% that of OFS!

Figure \ref{equ:7} gives the box and whisker plot for recorded
execution times of two feedback schedulers. It is clear that the
neural feedback scheduler induces much smaller computational
overhead than the optimal feedback scheduler. Thanks to the simple
form of the cost functions employed in the simulations, i.e.,
(\ref{equ:10}), the time overhead of the optimal feedback scheduler
is not very big. Intuitively, as the form of the cost functions
becomes more complex, the overhead of OFS will increase, but the
overhead of NFS will not be substantially affected.

\section{Conclusion}

As a fast and intelligent feedback scheduling scheme, the neural
feedback scheduling has been proposed in this paper for real-time
control tasks. It fully exploits the offline solutions for the
optimal feedback scheduling problem, which are offered by
mathematical optimization algorithms. With the proposed approach,
almost optimal QoC can be achieved. Meanwhile, compared to optimal
feedback scheduling, it can significantly reduce the runtime
overhead, which is particularly beneficial to embedded control
systems that operate in resource-constrained and dynamic
environments. The proposed approach does not rely on any specific
forms of the control cost functions, making it widely applicable. In
addition, the use of neural networks potentially enhances the
adaptability, robustness, and fault-tolerance of the feedback
scheduler.

\section*{Acknowledgments}
This work is partially supported by the China Postdoctoral Science
Foundation under grant number 20070420232, the Australian Research
Council (ARC) under Discovery Projects grant number DP0559111, the
Australian Government's Department of Education, Science and
Training (DEST) under International Science Linkages (ISL) grant number
CH070083, and the Natural Science Foundation of China under grant
number 60774060.


\end{document}